# Distribution of Traction Forces and Intracellular Markers Associated with Shape Changes During Amoeboid Cell Migration

Juan C. Lasheras[1,2,*], Baldomero Alonso-Latorre[1], Ruedi Meili[3],
Effie Bastounis[2], Juan C. del Álamo[1], Richard A. Firtel[3]

[1]*Department of Mechanical and Aerospace Engineering,*
[2]*Department of Bioengineering,*
[3]*Section of Cell and Developmental Biology,*
*University of California, San Diego. La Jolla CA 92093-0411*

During migration, amoeboid cells perform a cycle of quasi-periodic repetitive events (*motility cycle*). The cell length and the strain energy exchanged with the substrate oscillate in time with an average frequency, $f$, on top of which are imposed smaller random fluctuations. The fact that a considerable portion of the changes in cell shape are due to periodic repetitive events enables the use of conditional statistics methods to analyze the network of biochemical processes involved in cell motility. Taking advantage of this cyclic nature, we apply Principal Component Analysis (PCA) and phase-average statistics to analyze the dominant modes of shape change and their association to the activity and localization of molecular motors. We analyze time-lapse measurements of cell shape, traction forces and fluorescence from green fluorescent protein (GFP) reporters for F-actin in *Dictyostelium* cells undergoing guided chemotactic migration. Using wild-type cells (*wt*) as reference, we investigated the contractile and actin crosslinking functions of Myosin II by studying Myosin II heavy chain null mutant cells (*mhcA-*) and Myosin II essential light chain null cells (*mlcE-*). We found that the mechanical cycle of generation of traction stresses and cell shape changes remains remarkably similar for all cell lines, although these shape changes are implemented at a slower pace in Myosin II null mutants, probably due to a reduced control on the spatial organization of the traction stresses. We found that *wt*, *mlcE-* and *mhcA-* cells utilize similar modes of shape changes during their motility cycle. The number of these dominant modes of shape changes is surprisingly few, with only three modes accounting for almost 70% of the variance in all cases. These principal shape modes are dilation/elongation, bending, and bulging of the front/back. The first mode is associated with forward pseudopod protrusion, while the second mode, resulting from sideways protrusion/retraction, is associated to lateral asymmetries in both the cell traction forces and the distribution of F-actin, and is significantly less important in *mhcA-* cells.

*Corresponding author: lasheras@ucsd.edu

## INTRODUCTION

Amoeboid motility plays an important role in many physiological processes such as embryonic development, tissue renewal and the functioning of the immune system. It is also very important in pathological processes such as the metastatic spreading of some cancers (1). Directed cell motion may be triggered by several types of external stimuli. This migration process is controlled by a complex network of signaling biochemical pathways, which continuously drive the remodeling of the cytoskeleton of the cell and its adhesions to the extracellular matrix (ECM). The production and spatiotemporal organization of the traction forces exerted by the cell during migration is determined by the orchestrated interactions of actin-directed motors, F-actin regulation and crosslinking, motor protein-mediated contractility, and cell adhesions.

Despite an increasingly detailed knowledge of the biochemistry of the cytoskeleton and the pathways that regulate the cell remodeling during migration, a better understanding the spatiotemporal integration of these biochemical processes into specific events is still needed. In particular, the precise mechanisms whereby each stage of the motility cycle is related to specific biochemical signaling events are not yet clear.

Regardless of the complexity of the underlying biochemical pathways, the mechanical implementation of cell motility consists of a relatively simple cyclic succession of mechanical events: leading-edge protrusions, formation of new adhesions near the front, cell contraction, release of the adhesions and retraction from the rear (2) (Figure 1). Both leukocytes and *Dyctiostelium discoideum* in amoeboid form exhibit the above *motility cycle* with successive extension and retraction of pseudopods resulting in periodic temporal oscillations of the cell's length (3, 4), as well as the traction forces transmitted to the substrate (4–6).

Our method consists of simultaneously measuring the spatial and temporal changes in the distribution of fluorescently tagged cytoskeletal (or signaling) proteins and the distribution of traction forces that mediate each stage of the cell *motility cycle*, all the time also monitoring the changes in cell shape. We take advantage of the cyclic nature of amoeboid motility to apply conditional statistics and Principal Component Analysis (PCA) to connect specific biochemical processes to each of the physical events in the *motility cycle* by comparing *wt Dyctiostelium discoideum* amoeboid cells and mutant cell lines lacking cytoskeletal and regulatory molecules with known or suspected involvement in motility. This quantitative analysis will provide the necessary building blocks to begin constructing the complex network of biochemical processes controlling cell migration.

## ANALYSIS OF MIGRATION TRAJECTORIES

During chemotaxis, the cells' outlines were determined from differential interference contrast (DIC) images captured using a 40x lens at 2-sec intervals for all cell lines. Image processing was performed with MATLAB (Mathworks Inc, Natick, MA). This procedure is schematically shown in Figure 2. Figure 2 (a) shows the raw DIC image of the cell under study. We first remove the static imperfections of the image (b), and then enhance the edges of the cell by taking the natural logarithm of the absolute value of the intensity field (c). The probability density function $H(I)$ of the resulting intensity distribution $I(x,y)$ is shown in (d). The red line is a polynomial fit to this pdf; and the cyan dot indicates the location of the global maximum of $H(I)$, $I_0$. The green dots mark the position of $I_1$ and $I_2$, which define the region of the pdf whose centroid provides the instantaneous threshold $I_{thr}$ (yellow square). The white line in Figure 2 (e) shows the contour of the regions with values higher than $I_{thr}$ (this contour is displayed overlayed over the intensity image in panel b). We further perform a dilation/erosion with an element of size $S_0 = 6$ pixels as shown in (f). The final determination of the cell contour is obtained by applying an additional dilation/erosion operation with an element of size $S_1$ ($S_1$ is determined applying a semi-empirical expression relating the area covered by the detected features and the area which the contour of the cell should cover).

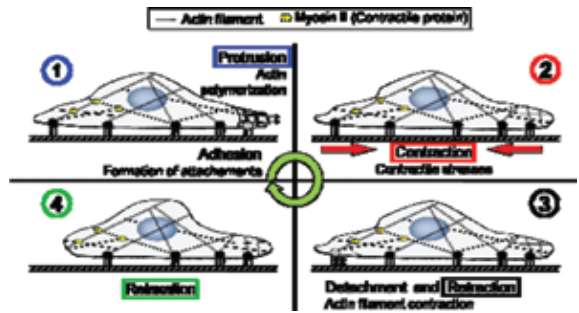

FIGURE 1
Sketch of the different stages of the *motility cycle* of an amoeboid cell.

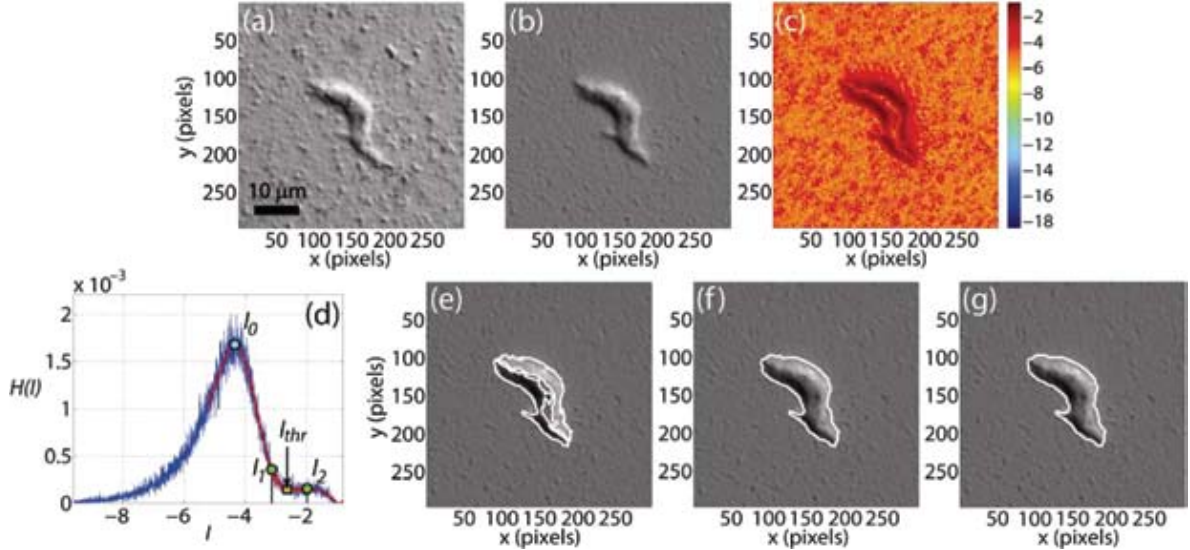

FIGURE 2
Cell contour detection algorithm.

## QUANTITATIVE EVIDENCE OF A FORCE REGULATED MOTILITY CYCLE

Amoeboid cells migrate undergoing a limited set of shape changes that compose the *motility cycle*: protrusion of a pseudopod at the front through actin polymerization, attachment to the substrate, contraction of the cell body, and detachment and retraction of the posterior part of the cell. This series of steps results in periodic variations of the cell length. During protrusion of the pseudopod the cell length increases while it decreases during the retraction of the rear. In addition, the strain or elastic energy exerted by the cell on the elastic substrate (7) also varies periodically and in-phase with the cell length, as shown in Figure 3a and extensively discussed in (4, 8).

The existence and degree of periodicity of a force-regulated *motility cycle* is demonstrated in both the auto and cross correlation of cell length ($L$) and strain energy ($U_s$) shown in Figure 3a-b for a typical *wt* cell. We observe that both $L$ and $U_s$ fluctuate in a cyclic fashion and are highly correlated. The autocorrelation of $U_s$, $R_{UsUs}$, and the cross-correlation between $L$ and $U_s$, $R_{LUs}$, also show a marked degree of periodicity. The period of the motility cycle, $T$, is determined as the time lapse between peaks. The magnitudes of the peaks for both $R_{UsUs}$ and $R_{LUs}$ are very similar, meaning that $U_s$ is as correlated with itself as it is with $L$, and evidence of the correlation between the variations in the cell length and the level of stresses transmitted to the substrate, i.e., between the kinematics and the dynamics of the cell. Similar results are obtained when examining *mhcA-* and *mlcE-* cells. In fact, the probability density functions of the correlation coefficient between $U_s$ and $L$, $r_{LUs}$, for all the three cell lines under study, shown in Figure 3c, indicate that they are strongly correlated in all our experiments. The percentage of cells showing a correlation coefficient $r_{UsL}$ larger than 0.5 is 33% for *wt*, 49% for *mhcA-* and 55% for *mlcE-*.

## THE VELOCITY OF MIGRATION IS DETERMINED BY THE PERIOD OF THE MOTILITY CYCLE

Our measurements have shown that the velocity of *Dictyostelium* cells chemotaxing on flat surfaces is determined by the rate at which the cells are able to

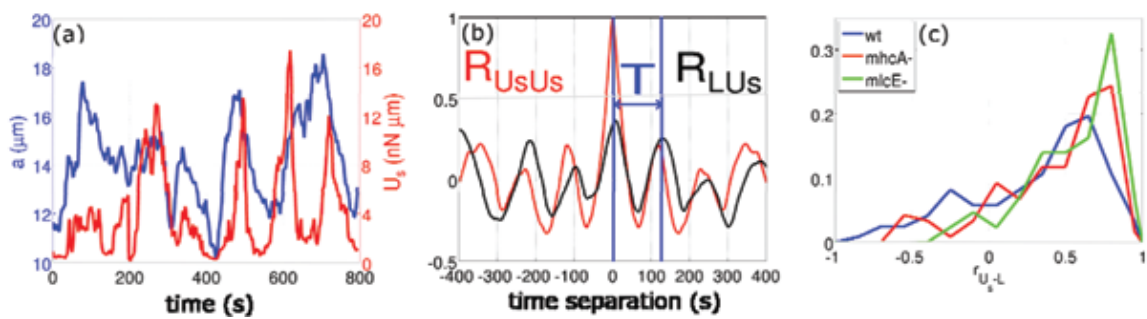

FIGURE 3
(a) Example of the temporal evolution of the major cell semiaxis, *a* (blue) and the strain energy, $U_s$ (red) for a *wt* cell. (b) Auto-correlation of the strain energy, $R_{UsUs}$ (red); and cross-correlation between cell length and strain energy, $R_{LUs}$ (black); as a function of the time separation. (c) Histogram of the correlation coefficient between the strain energy $U_s$ and the length of the cell $L$ for *wt* (blue, N = 31 cells), *mhcA*- (red, N = 27 cells) and *mlcE*- (green, N = 14 cells) cells.

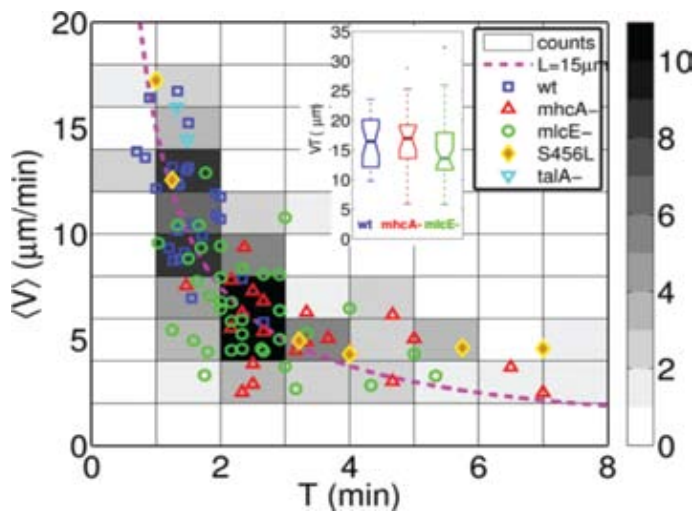

FIGURE 4
Scatter plot of the average velocity of N = 91 chemotaxing *Dictyostelium* cells versus the period of their motility cycle. The data points come from five different cell lines: N = 25 *wt* cells (blue squares), N = 21 *mhcA*- cells (red triangles), N = 37 *mlcE*- cells (green circles), N = 6 *S456L* cells (orange diamonds), and N = 2 *talA*- cells (cyan triangles). The dashed magenta hyperbola ($V = L/T$) is a least square fit to the data, yielding $L = 15.1~\mu m$. The $V - T$ plane has been divided into tiles that have been colored according to the number of cells whose speed and motility period lie within each tile. Darker tiles contain more cells, as indicated in the color. The inset shows box plots of the product $VT$ for *wt*, *mhcA*- and *mlcE*- cells.

repeat their *motility cycle* (Figure 4). The relationship between the average migration velocity of a cell (*V*) and the period of its strain energy cycle (*T*) is well approximated by the hyperbola $VT = L_0$, where $L_0$ is a constant with units of length (4). The correlation coefficient between *V* and $1/T$ is R = 0.75.

The fact that, regardless of the speed of migration, a cell advances on the average a fixed length per cycle suggests that although there are many interconnected molecular regulatory loops, the complex mechano-chemical system must settle into a relatively simple quasi-periodic cycle.

## PHASE-AVERAGE ANALYIS OF THE MOTILITY CYCLE

We have developed a method to systematically obtain statistically-significant information of the spatial biochemical and mechanical organization of the cell at each stage during their migration cycle. Our method consists of dividing the motility cycle into canonical stages (or phases) and then performing phase-averaged measurements cell shape, distribution of traction forces, and molecular concentrations of fluorescent protein reporters. The resulting unified representation of the motility cycle enables us to directly evaluate in large cell populations the mechanical function of a particular protein by comparing cell lines in a statistically robust manner, instead of comparing individual cells subjectively designated as representative. In the following paragraphs, we apply this statistical technique to obtain quantitative comparison of the spatial organization and magnitude of the traction forces of *wt*, *mlcE-* and *mhcA-* cells.

We divided the motility cycle into four stages (bins), as shown in Figure 5: 1) *Protrusion*, marked in black in the figure, and defined as the fraction of time during which the cell length is increasing; 2) *Contraction*, marked in red and defined as the fraction of the motility cycle for which the length of the cell is at a local maximum (our force measurements described in the previous section have shown that this coincides with the time when the cell is exerting maximum strain energy); 3) *Retraction*, marked in green and defined as the fraction of the cycle during which the cell length decreases; and 4) *Relaxation*, marked in blue and defined at the fraction of the cycle when the cell length is at a minimum (our force measurements have shown to coincide with the fraction of the cycle of minimum strain energy). Each measurement is then sorted out into one of the above four bins according to the stage at which the cell is in its motility cycle.

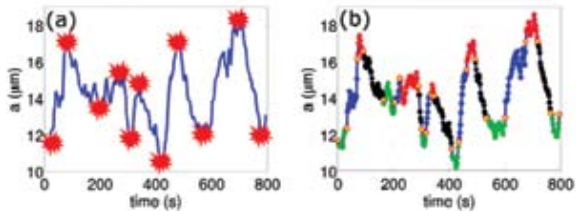

FIGURE 5
Representation of the steps of the algorithm used to calculate the phase averages. (a) After the major semiaxis of the cell, *a(t)*, is recorded for every frame the peaks and valleys of each time history are identified. Panel (a) shows the time evolution of the semiaxis of a *wt* cell (blue line) and the determined location of the peaks and valleys of that time evolution. (b) A computer algorithm automatically selects the 4 phases. Panel (c) shows the output of the automatic dissection of the *motility cycle* into stages: protrusion (blue), contraction (red), retraction (black) and relaxation (green).

To perform the statistical analysis of each bin (stage), we have also developed a suitable methodology that accounts for the fact that the shape and the orientation of the cells are constantly changing and they are not the same for all cells in each bin. We first compute for each cell measurement the location of the center of mass (centroid) of the cell and the orientation and length of its major and minor axes of inertia. We scale at each time the dimensions of the cell with the instantaneous value of the major semi axis length, a. Each measurement is then represented in this cell-based dimensionless coordinate system by placing the centroid $(x_c,y_c)$ of the cell at the origin of the coordinate system and rotating the cell to have its major moment of inertia always coinciding with the horizontal axis. Note that in this cell-based coordinate system the cell is always approximately located between $-1 < x/a < 1$ [4].

Figure 6 demonstrates that the above described cycle-sorting algorithm is able to statistically dissect the motility cycle of *Dictyostelium* cells into the succession of leading-edge protrusions, formation of new adhesions near the front, cell contraction, release of the rear adhesions and retraction that has been described phenomenologically elsewhere [2]. Figure 6 shows phase-averaged maps of the fluxes (gain/loss) of cell area that occur during migration. The red and blue areas in the figure indicate the locations where area is being added (red) or depleted (blue) respectively during each of the stages of the motility cycle.

Notice that if the cell moves as an imaginary rigid object gliding over the substrate, the area gain at the front of the cell would always equal the area loss at the back in all panels. However, our measurements show a quite different behavior. During the protrusion stage, new material is being added at the leading edge while little area is lost at the back of the cell. Conversely, during retraction there is significant area loss at the back and virtually no area gain at the front. These results are consistent with the analysis of Soll and colleagues [3]. The emergence of these clearly distinguishable stages from the area flux analysis strongly suggests that our cycle-sorting algorithm is indeed able to separate quantitatively the different stages of the motility cycle of *Dictyostelium* cells.

The success of this approach to measure and describe the averaged stereotypical cell behavior allows us to compare the biophysical characteristics of each stage in different mutant strains. We observe that the area fluxes are lower in *mhcA-* and *mlcE-* cells than in wt cells during the four stages of the motility cycle, indicating that myosin II plays an important role in all stages, including protrusion. Apart from the differences in absolute value, the average spatial distribution of the area fluxes of *mhcA*⁻ and *mlcE*⁻ cells is similar to *wt* in all the stages, suggesting that even though the time required to complete the cycle increases substantially, the mechanical implementation of the motility cycle does not change fundamentally despite the lack of actin-myosin contractility.

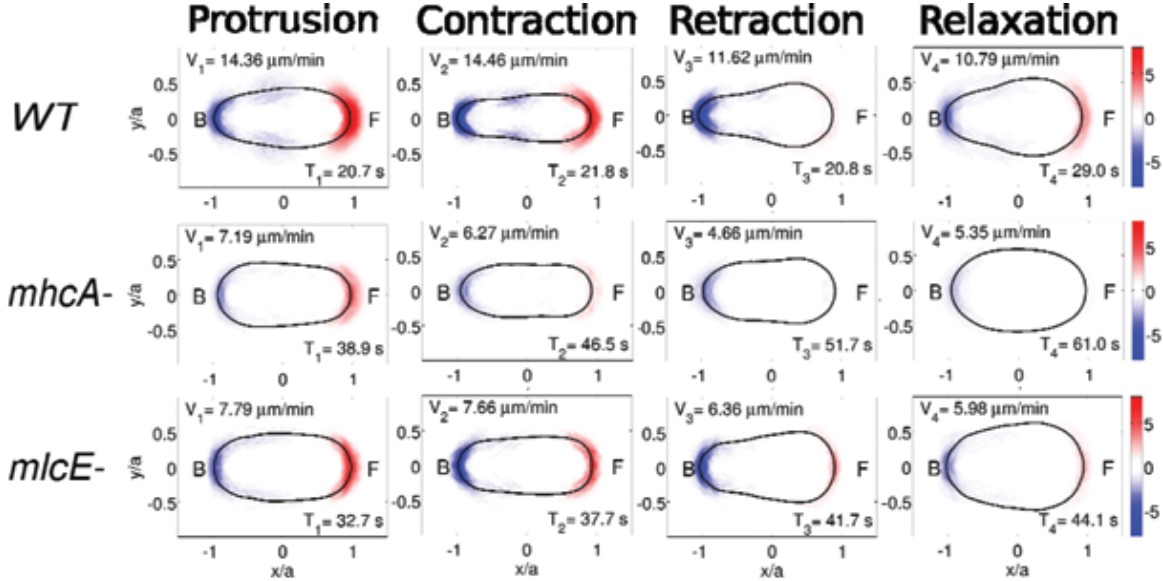

FIGURE 6
Phase-averaged cell shape and area fluxes during the four stereotypical stages defined in Figure 5b for *wt* (top row, N = 31 cells), *mhcA*⁻ (middle row, N = 27 cells) and *mlcE*⁻ (bottom row, N = 24 cells) cells. The contour maps show the average rate of change of cell area, computed in a reference frame rotated to coincide with the instantaneous principal axes of the cells and scaled with the length of their instantaneous major axis, *a*. The colors indicate the magnitude of the area flux in $\mu m^2/s$. The black contours show the average shape of the cells. The front (F) of the cell corresponds to x > 0 and the back (B) corresponds to x < 0. The legend in each panel shows the average duration of each stage $T_1 - T_4$.

## PRINCIPAL COMPONENT ANALYSIS (PCA) OF THE SHAPE, TRACTION FORCE, AND CHEMICAL INTRACELLULAR MARKERS

In spite of the complex myriad of biochemical processes involved, cell migration is the result of the quasi-periodic repetition of a reduced set of steps (and shape changes). Therefore, it is fundamental to identify not only those shape changes that are repeated cyclically, but also the biochemical and mechanical processes involved in the generation of each these shape changes. For this purpose, we have implemented Principal Component Analysis (PCA) (9), which enables us to identify for each cell line the dominant shape patterns that occur during the migration of a cell together with the weight factors (variance) that determine the relative contribution of each mode to cell shape at each instant of time. We consider the shape of the cell as a two dimensional scalar field allowing us to relate it to the distributions of measured traction forces or to the concentration of biochemical markers (F-actin) associated with each of the principal shape modes.

We found that the shape modes obtained from PCA are highly reproducible from cell to cell. Four out of the five most-representative shape mode patterns of each cell were common for all the cells studied. Furthermore, these four common, most representative modes in the PCA are enough to account for three quarters of the observed shape variance in migrating *Dictyostelium* cells (75% ± 4% for N = 23 *wt* cells, 75% ± 4% for N = 22 *mhcA*⁻ cells and 76% ± 3% for N = 15 *mlcE*⁻ cells). These results indicate that the dynamics of the shape changes in all migrating cells can be approximated using a small number of degrees of freedom, in agreement with previous PCA of the cell contour performed in migrating *wt Dictyostelium* cells and keratocytes (10, 11). Figure 7 shows the four most relevant shape mode patterns and their associated traction forces for a *wt* cell. The modes for *mhcA-* and *mlcE-* cells are similar (Figure 8). Figures 7 and 8 also contain histograms of the instantaneous values of the weight factors of each PCA mode. Positive/negative values of those weight factors correspond to the $(+)$/$(-)$ configuration of each mode.

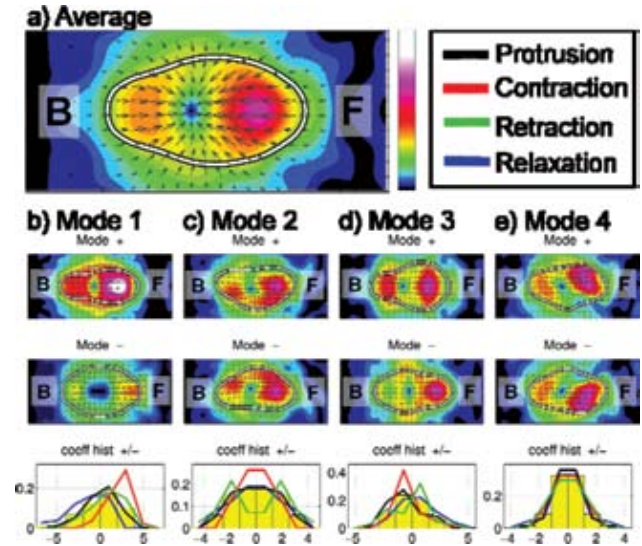

FIGURE 7
PCA of the shape and traction forces of a migrating *wt Dictyostelium* cell. (a): average shape and traction forces. The white contour indicates the contour of the cell. The color contours indicate the magnitude of the stresses and the arrows indicate their direction. (b)-(e): four most dominant shape modes patterns for positive (top) and negative (bottom) values of the weight factor. The bar plots in the lower panels show the distribution of instantaneous weight factors for this cell during the whole time history of the cell (yellow bars) and during each stage of the *motility cycle* (color curves). For all panels "F" and "B" indicate the front and the back of the cell respectively.

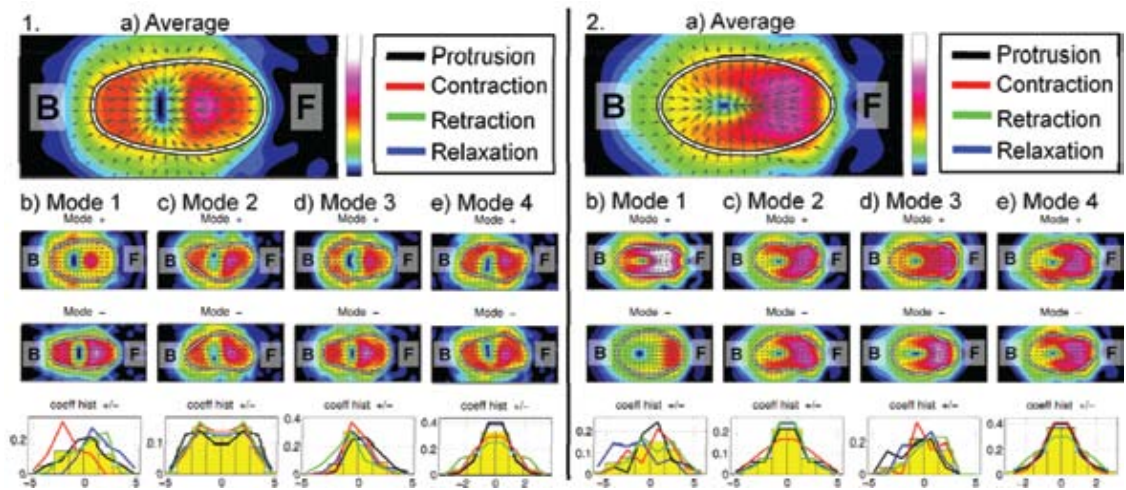

FIGURE 8
PCA of the shape and traction forces of a migrating *mhcA- (1)* and *mlcE- (2) Dictyostelium* cell.

In most of the cell lines studied, the most important shape mode in terms of variance is Mode 1, which accounts for the changes in the aspect ratio of cell shape due to the periodic oscillations in cell length. As one would expect, the weight factors are predominantly positive during contraction, when the cell reaches its maximum length. Conversely, during relaxation, when the cell is roundest, the weight factors are predominantly negative. Note that the forces associated with the (+) shape pattern of Mode 1 (contraction phase) are high, whereas the forces associated with the (−) one (relaxation phase) are very low. Mode 2 is the second most statistically relevant shape mode. It is associated with the bending of the cell (a half-moon shape) and lateral asymmetries in the traction forces. This mode represents the events of lateral protrusion and retraction of pseudopods. Mode 3 represents a non-symmetric bulging and contraction of the front and back. It is important to notice that the relative contribution of the positive and negative modes differ between protrusion and retraction. The (−) mode is prominent during protrusion, when the cell is extending a frontal pseudopod and the (+) mode during retraction, when the cell is retracting its posterior part. The traction stresses associated to the (+) mode are higher in the front than in the back of the cell. In the (−) mode the increment of the force at the front of the cell is balanced by a concentric cortical/peripheral distribution of forces, which together with the circular cell contour at the rear of the cell is consistent with a possible hydrostatic loading (pressure) of the cytoskeleton. The (+) configuration of Mode 3 is characterized by a more concentrated spot of traction stresses at the back of the cell, especially for *wt* cells.

Figure 9 shows the same four principal shape modes for a *wt* cell, but in this case we have added the data obtained using the fluorescently tagged reporter LimE-Δcoil-GFP (12) that binds to F-actin. As shown, the contour maps correspond to the associated fluorescence intensity of LimE-Δcoil-GFP [10]. Figure 9 also contains histograms of the instantaneous values of weight factors of each PCA mode for the duration of each experiment.

**Relevance of the PCA modes**

A comparison of the importance of the four most relevant PCA modes during the stages of the motility cycle of three cell lines of *Dictyostelium* cells (*wt*, *mhcA-* and *mlcE-*) is shown in Figure 10. The data indicates that Modes 2 and 4 are substantially less important in *mhcA-* cells than in the other two strains, indicating that purely frontal protrusion is more dominant over sideways protrusion for

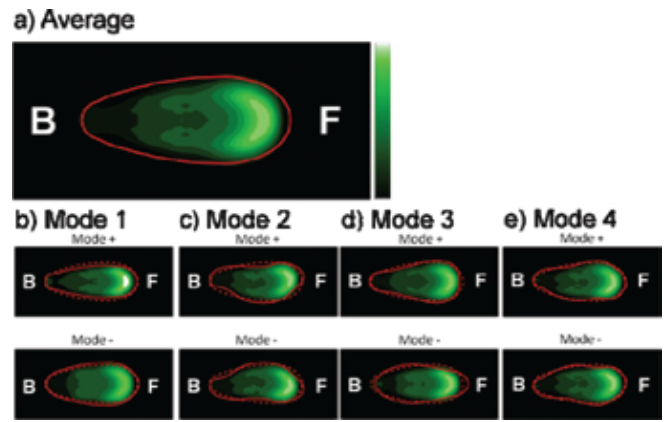

FIGURE 9
PCA of the shape and fluorescence intensity produced by the F-actin binding protein LimE-GFP for a migrating *wt Dictyostelium* cell. The top panel shows the average data. The red contour indicates the contour of the cell. The color contours indicate the intensity of the LimE-GFP fluorescence. The lower panels show the shape and fluorescence patterns determined by the four most important modes, similar to Figure 8. For all panels "F" and "B" indicate the front and the back of the cell respectively.

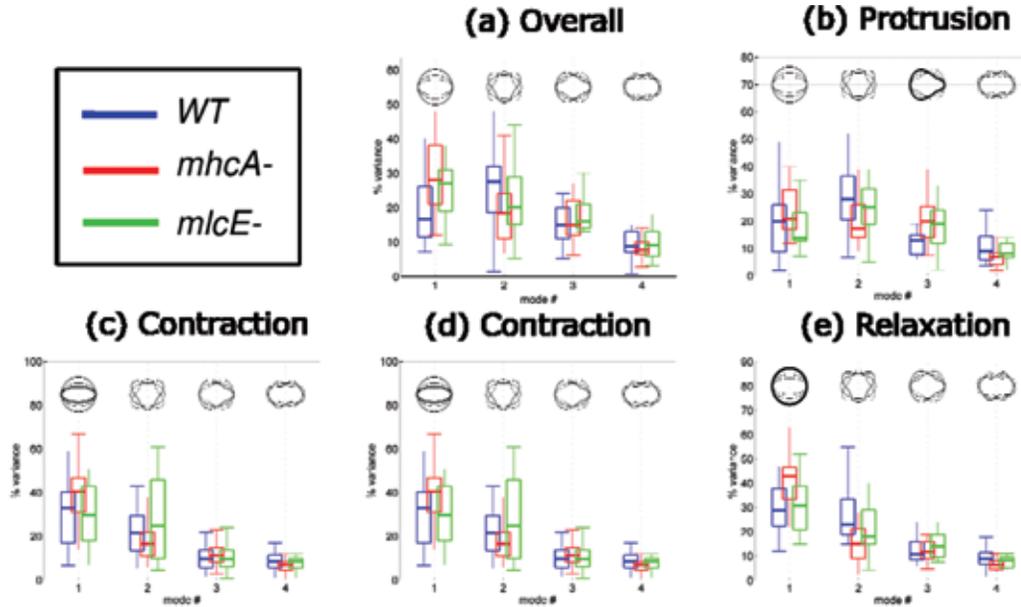

FIGURE 10
Box plots of the percentage of cell shape variance associated to each of the four principal modes for *wt* (blue, N = 23 cells), *mhcA*- (red, N = 22 cells) and *mlcE*- (green, N = 15). (a) during the overall motion of the cells; (b) during protrusion; (c) during contraction; (d) during retraction.

*mhcA*- cells, whereas the opposite holds for *wt* and *mlcE*-. Sideways protrusion of pseudopods are associated with local enhancements of lateral tension and cortical F-actin at one side of the cell, which is required to balance the internal shear produced on the actin cytoskeleton. Otherwise the protruding pseudopod could not be mechanically stable. The implication is that *mhcA*- cells are less effective in controlling their lateral tension and therefore have impaired sideways protrusion. This defect may explain the reduced frequency of pseudopod protrusion in *mhcA*- cells, which was observed by Wessels *et al* (3). Consistent with these ideas, Fukui *et al* (13) reported reduced stability of protruding pseudopodia in *mhcA*- cells subjected to centrifugal forces.

## ACKNOWLEDGMENTS

This work was supported in part by 1R01 HL0805518 NIH , BRP081804F NIH. Juan C. del Álamo was partially supported by a Fulbright Scholarship from the Spanish Ministry of Education and the US Departent of State.

## NOMENCLATURE

| | |
|---|---|
| DIC | Differential interference contrast |
| ECM | Extra-cellular matrix |
| GFP | Green fluorescent protein |
| $L$ | Length of the major moment of inertia of the cell (μm) |
| *mhcA*- | Myosin II heavy chain null mutant cells |
| *mlcE*- | Myosin II essential light chain null cells |
| PCA | Principal Component Anaysis |
| pdf | Probability density function |
| $R_{UsUs}$ | Autocorrelation of $U_s$ |
| $R_{LUs}$ | Cross-correlation between $L$ and $U_s$ |
| $r_{LUs}$ | Correlation coefficient between $U_s$ and $L$ |
| $T$ | Period of the motility cycle (min) |
| $U_s$ | Strain energy exchanged between the cell and the substrate (pN μm) |
| *wt* | wild-type cells. |